\documentclass[12pt,prd,nofootinbib,superscriptaddress]{revtex4}
\usepackage{graphicx}				
\usepackage{amsmath}
\usepackage{amsfonts}

\begin{document}
\title{Beyond information: A bit of meaning.}
\author{Olaf DREYER}
\affiliation{Dipartimento di Fisica, Universit\`a di Roma ``La Sapienza"\\
and Sez.~Roma1 INFN, P.le A. Moro 2, 00185 Roma, Italy}

\begin{abstract}  Is our world just information?  We argue that our current notion of information has one serious shortcoming:  It is quite literally meaningless.  We suggest a meaningful extension of the notion of information that is dynamic, internal, approximate, contains an element of randomness,  and is layered.  This new notion of information derives from the interactions of material objects.  Our answer to the essay question then is Bit from It or,  more appropriately,  Bit$^{++}$ from It.  We discuss how our new notion of information sheds light on the measurement problem in quantum mechanics and how it can be applied in philosophy and computer science.
\end{abstract}

\maketitle

\tableofcontents

\newpage

\section{Sphingidae can suck it}\label{sec:intro}
\begin{quote}
I have just received such a Box full from Mr Bateman with the astounding Angr¾cum sesquipedalia with a nectary a foot long -- Good Heavens what insect can suck it [\ldots]\\
\emph{Letter from C.~R.~Darwin to J.~D.~Hooker,  25. January 1862\cite{darwinletter}}
\end{quote}

\begin{figure}[htbp]
\begin{center}
\includegraphics[width=12cm]{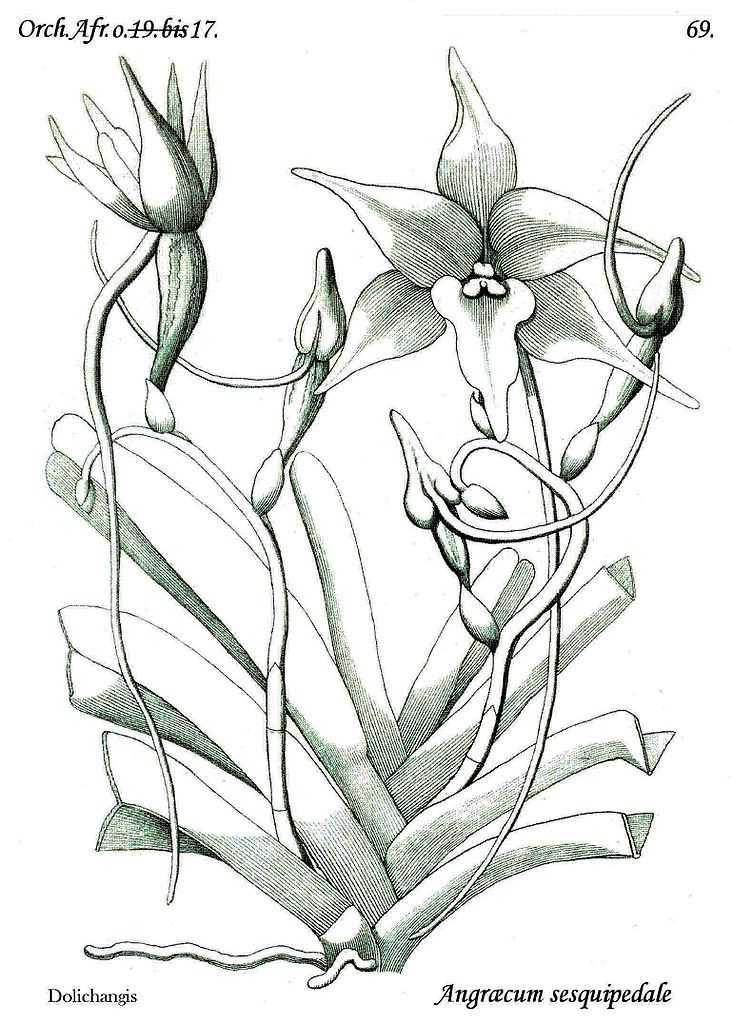}
\caption{Angraecum sesquipedale, or Darwin's orchid.  The remarkable feature of this orchid is its very long spur which extends from behind the flower and grows to a length of over 30 cm.  The nectar is found on the bottom of this spur.  Both Darwin and Wallace predicted that there should be an insect with a proboscis that is long enough so that it can get to the nectar.  The Sphinx moth was found in 1903 in Madagaskar. \cite{figureorchid}\label{fig:orchid}}
\end{center}
\end{figure}

In 1862 Darwin was sent a box containing samples of angraecum sesquipedale,  an orchid that possesses a spur that is over 30 cm long.  This spur is a part of the flower that grows behind the head of the flower and contains the nectar at its bottom (see figure \ref{fig:orchid}).  Immediately after seeing the orchid Darwin conjectured that there should be an insect with a proboscis\footnote{A proboscis is the elongated snout of an insect that allows it to reach parts deep inside a flower.}   that is long enough to reach the bottom of the orchid's spur\footnote{Wallace went further and suggested that the insect in question was in fact a moth.}.  The insect in question was not found until 1903.  It is the sphinx moth,  or sphingidae,  from Madagascar and it does indeed have an elongated proboscis that can reach the bottom of the sphinx orchid.

Without ever having seen the moth Darwin was able to infer its existence by looking at the orchid.  The orchid is able to do this through its shape.  Anything that wants to get at the nectar at the bottom of the spur has to be very particular.  If the tool for extracting the nectar is too short it can't reach the nectar.  If it is too wide it can't enter the spur.  It is the way the spur interacts with other objects like it that allowed Darwin to infer details about the sphinx moth.  The orchid is a representation of the sphinx moth because of how it interacts with other objects like it (proboscises,  beaks, ... ).

This is the first and most important characteristic of our new notion of information.  The orchid is a representation of the moth.  The orchid acquires this meaning by the way it interacts with other objects.  This is a form of information that includes meaning through interaction. 

Let us remark on two other characteristics of this kind of information.  The first thing is that the representation is approximate.  It captures some qualities of the moth but not others.  The shape of the orchid gives no indication of the color of the moth for example.  

The other important characteristic is that the representation of the moth by the orchid is completely internal to the orchid.  Nothing else other than the orchid is needed to infer the moth. 

\section{A simple example}
\begin{quote}
We are so accustomed to this rigidity property that we do not accept its almost miraculous nature,  that it is an ``emergent property" not contained in the simple laws of physics,  although it is a consequence of them.\\
\emph{P.~W.~Anderson}  Basic notions of condensed matter physics\cite{anderson}.
\end{quote}

\begin{quote}
I refute it thus.\\
\emph{S.~Johnson}\footnote{To refute Bishop Berkeley's idealism Samuel Johnson kicked a stone while exclaiming ``I refute it thus" (for more details see \cite{boswell}).}  
\end{quote}

Let us look at an easier example of this kind of information.  Possibly the simplest example is that of a solid representing a position.  If we want to represent a position we can just put a solid at the position that we want to represent.  In the introduction we have seen three characteristics of the kind of information that we are talking about here:  meaning arises through interaction,  it is dynamic,  and it is internal.  We can immediately see all these three characteristics in this simple example.  The solid represents a position by being at the position.  The way to infer the position is to take another solid and probe the space.  If one bumps into the solid one has found the position.  This process is dynamic because it requires the interaction of the two solids.  It is also internal.  This representation of a position does not require an external dictionary that tells us what the representation means.  The meaning of the solid comes from how it interacts with other solids.  An external representation would look like this:
\begin{equation}
(x,y,z)
\end{equation}
This representation is external because the three numbers alone do not tell us enough about the position that we want to describe.  What are the axis?  What are the units?  But things are even worse than that.  The three numbers could represent many other things other than a position.  They could be the Euler angles of a rotation,  the GDP figures of Great Britain in the last three years,  etc.  To properly understand the three numbers additional,  external,  information needs to be given.  This is not the case with the representation of the position by a solid.  The solid is all that is needed to infer the position.  

\begin{figure}[htbp]
\begin{center}
\includegraphics[width=15cm]{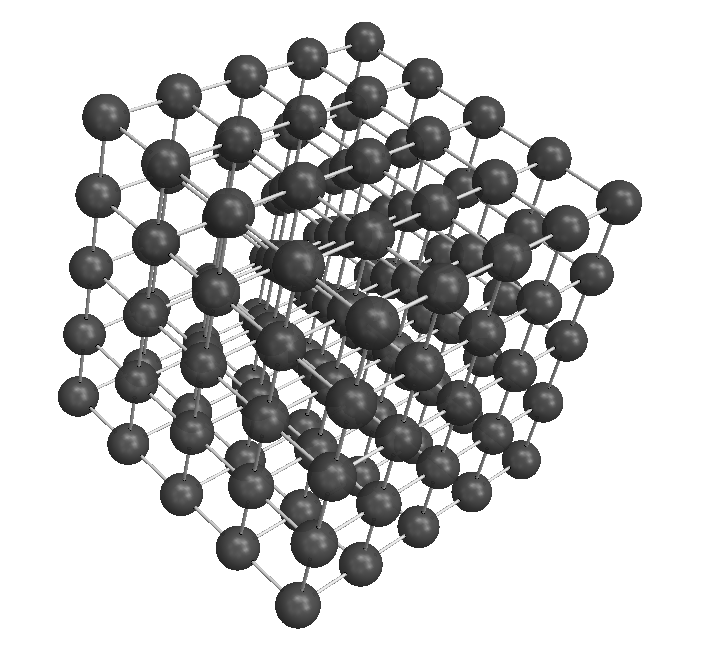}
\caption{The lattice is an emergent property of a solid.  A consequence of the lattice is the rigidity of the solid:  when you push the solid it pushes back in an attempt to maintain the lattice.  The lattice also breaks the symmetry of space.  The molecules that make up the solid are at certain non-random positions.  This breaking of the symmetry introduces an inevitable element of randomness in the process of creating the solid.  \label{fig:lattice}}
\end{center}
\end{figure}

The example of the position allows us to give two more characteristics of our kind of representation:  The existence of layers and the appearance of randomness.  

In our representation of the position it is important that one solid prevents another solid from moving any further when it bumps into it.  This rigidity is such a common feature of our world that we hardly ever pause to consider how remarkable it is.  It is a consequence of the lattice of molecules that make up the solid.  The solid wants to maintain this lattice.  This means that a force on one molecule will move the whole lattice.  It is this resistance that we feel when we bump into a solid and it is a feature that the solid has but the molecules do not.  We can repeatedly bump into a solid to check its position while we can not do the same with a molecule.  One bump and the position of the molecule will have changed so much that we can not use the molecule to represent that position.  

The lattice and the resulting rigidity are true emergent properties of a very large number of molecules.  There are thus two layers of objects with different properties:  the layer of solids and the layer of molecules.  The layer of solids has objects that have the property of rigidity that allows them to represent positions.  The layer of molecules does not have objects with this property.

When a solid together with its lattice forms a remarkable thing happens:  a particular position for the lattice gets chosen.  If we take a cubic lattice as an example we need just two things to describe the whole lattice,  the position of one molecule and the three vectors that give the edges of one basic cell.  We can built the whole lattice by putting molecules at the ends of the vectors that are obtained by adding all possible multiples of the basic vectors.  Before the solid is formed the system is completely symmetric.  There is no special position in space.  After the solid has formed the lattice is breaking this symmetry.  The position of the one molecule is special.  The choice of this particular position depends on small random fluctuations that are present when the solid forms.  There is an inevitable element of randomness involved when one goes from one level to the next.  

\section{The six characteristics of meaning}\label{sec:meaning}
Let us gather the different characteristics of information that we have seen so far.  The most important aspect is that the meaning of an object arises from the way it interacts with similar objects.  The physical restrictions of the orchid imply the moth and the rigidity of the solid gives its position.  An important corollary of this is that meaning is dynamic and not static.

\begin{table}[htbp]
\begin{tabular}{rcl}
\textbf{Interaction} & \ \ \ \ \ \ \ \ \ \ \ & Meaning arises from the interaction \\
  & & of objects of the same kind.\\
 & & \\
\textbf{Dynamic} & & Meaning is dynamic and not static.\\
 & & \\
\textbf{Internal} & & Meaning is internal and not external.\\
 & & \\
\textbf{Approximate} & & Meaning is approximate.\\
 & & \\ 
\textbf{Random} & & The process of emergence \\
 & & includes elements of chance. \\
 & & \\
\textbf{Layered} & & Meaning arises in layers. \\
 & & \\
\end{tabular}
\caption{The six characteristics of our new notion of information.\label{tab:meaning}}
\end{table}

The meaning is also internal.  No external encyclopedia is needed to tell us that the orchid represents the sphinx moth or that the solid represents a certain position.  The physical properties of the orchid and the solid tell us that.

Furthermore meaning that arises in this way is necessarily approximate.  We have seen this in the example of the orchid.  The form of the orchid gives information about the proboscis of the moth but not about its color.  

The example of the solid gave us two more characteristics:  randomness and layers.  The solid acquires its meaning through its rigidity which goes back to the lattice of molecules that is making up the solid.  Randomness enters the picture because the lattice breaks the translational symmetry of space.  When the solid is formed one choice for the position of the lattice has to be made.  This choice is determined by random fluctuations of the environment in the moment of creation.  Another example of this kind is given by the ground state of a large number of interacting spins.  In the ground state all spins point in the same direction.  This direction breaks the rotational symmetry of space and is again chosen by random fluctuations during the formation of the ground state\footnote{Just like in the case of the solid the spins want to maintain this direction.  In the case of the spins this tendency to resist change is called generalized rigidity (see \cite{anderson,chaikin} for more details).}.

The rigidity that is so important for our new notion of information appears only on the level of solids and not on the level of the molecules that make up the solid.  Furthermore,  the meaning of the solid only arises from interactions with other solids not with molecules.  We thus have a natural layered structure to our notion of information.  The layers are related to each other through a process of emergence.  Table \ref{tab:meaning} summerizes these six characteristics of information.  

\section{A Gedankenexperiment}
\begin{quote}
I, at any rate, am convinced that He does not throw dice.\\
\emph{A.~Einstein},  Letter to Max Born\cite{bornletters}.
\end{quote}

\begin{figure}[htbp]
\begin{center}
\includegraphics[width=12cm]{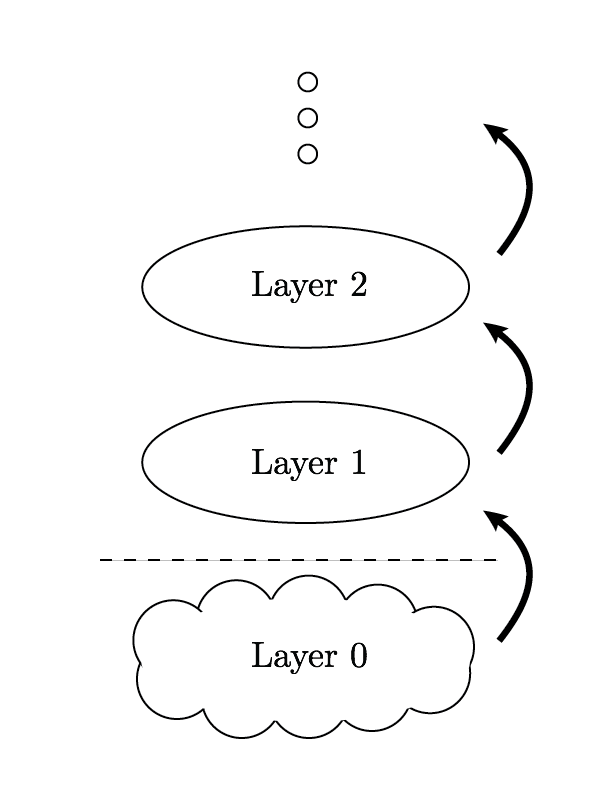}
\caption{The tower of layers.  The arrows indicate the direction of emergence.  We pose the following question:  How does layer 0 look like to someone who's lowest level meaningful objects are from layer 1? \label{fig:tower}}
\end{center}
\end{figure}

We are now in the position to ask an interesting question.  One of the characteristics of our new notion of information is that it is layered.  The meaning of objects arises from their interaction with objects in the same layer and different layers are related to each other through a process of emergence.  Now imagine the following situation:  Imagine that what we thought of as the lowest,  most basic, layer turns out not to be the lowest layer.  Instead,  what we thought were the most basic objects turn out to be emergent from objects on an even lower layer (figure \ref{fig:tower} depicts this situation).  How would that look to us?

The first thing that we will note is that these more basic objects (layer 0 objects,  see figure \ref{fig:tower}) will lack properties that we think of as fundamental.  Because they are not on the same level as the objects we thought were the most basic objects (layer 1 objects) they do not have the same set of properties.  Trying to interact with them as we do with other level 1 objects will lead to inconsistent and confusing results.  

We will try to assign to level 0 objects the same properties that we assign to level 1 objects.  We will do this by devising methods to create level 1 objects whose properties we can easily verify.  As we have seen in the previous sections this process introduces an irreducible element of randomness.  We will thus be faced with a situation where chance enters our description of nature in a fundamental way.

Is this just an academic Gedankenexperiment?  The quote at the beginning of this section indicates that we think it is not.  In fact we think that this is exactly what happened to us in the beginning of the last century when we discovered quantum mechanics.  We discovered objects that do not seem to have such fundamental properties as position.  We devised ways to create level 1 objects (like bubbles in a cloud chamber) that allowed us to assign (or measure) properties of these level 0 objects and we discovered randomness in the process.  We propose that some of the puzzling features of quantum mechanics can be understood with our new view of information.

\section{It from bit or Bit from it?}
\begin{quote}
42\\
\emph{Douglas Adams}, The hitchhiker's guide to the galaxy.
\end{quote}

\begin{quote}
Frequently the messages have \emph{meaning}; that is they refer to or are correlated according to some system with certain physical or conceptual entities. These \emph{semantic aspects of communication are irrelevant to the engineering problem}. The significant aspect is that the actual message is one selected from a set of possible messages.\\
\emph{C.~E.~Shannon}, A Mathematical Theory of Communication \cite{shannon} (emphasis added)
\end{quote}

\begin{quote}
The birth of information theory came with its ruthless sacrifice of meaning -- the very quality that gives information its value and its purpose.\\
\emph{J.~Gleick}, The Information \cite{gleick}
\end{quote}

If we have to decide between "It from bit" and "Bit from it" it is clear that we come down on the latter:  Bit from it.  What we have spent most of this essay on though is our conviction that the bit-part needs to be improved.  A bit is usually seen as the basic unit of information.  The question "It from bit" or "Bit from it" is then the question about what is more fundamental,  information or matter?  A number of people have suggested that information should be seen as the basis of our description of the world \cite{lloyd,vedral}.  Our analysis shows that there is something fundamentally wrong with this suggestion.  Naked bits require a dictionary that gives them meaning.  Such a dictionary is necessarily \emph{external} to the bits themselves and a description of the world that focuses solely on the bits will be incomplete.  We have shown that meaning can be internal but it requires us to give up the idea that our world is pure information.

The narrow view of information that was introduced by Shannon served us well during the fast evolution of computer technology in the last fifty years but we think that we are now running up against its limitations.  We have already hinted at how our new view of information can be used to see the measurement problem in quantum mechanics in a new light.  Other possible applications include philosophy and computer science itself.

One of the perennial problems in philosophy is the problem of consciousness.  One reason consciousness is puzzling is that there seems to be an infinite regression present.  It feels like there is someone observing the thoughts inside our head but then what about the thoughts of that someone?  D.~Dennett called this view of consciousness the Cartesian Theatre\cite{dennett}.  Our view of information might be able to break this infinite loop because the meaning of our objects is internal.  There is no outside observer required.  The thoughts acquire meaning through the way they interact with other thoughts.

Our view of information suggests that there should be a new paradigm of computation that we might call emergent computation.  This computation will consists of the dynamic evolution of emergent objects.  The six characteristics that we have outlined in section \ref{sec:meaning} will be present here.  In particular the computation will by necessity include random elements and be approximate.  Two properties not shared with our current model of computation.  The most important aspect of emergent computation will be that the meaning of the objects in the computation is completely internal.


\begin{thebibliography}{WW}
\bibitem{darwinletter}  The whole letter can be found at the Darwin Correspondence Project (www.darwinproject.ac.uk).  The number of the letter is 3411.
\bibitem{figureorchid}  The figure is taken from the wikipedia article on angraecum sesquipedale and is in the public domain.
\bibitem{anderson} P.~W.~Anderson,  \emph{Basic notions of condensed matter physics},  The Benjamin/Cummings Publishing Company,  Inc. (1984).
\bibitem{boswell} J.~Boswell,  \emph{The life of Samuel Johnson},  Penguin Classics (2008). 
\bibitem{chaikin} P.~M.~Chaikin and T.~C.~Lubensky,  \emph{Principles of condensed matter physics},  Cambridge University Press (1995).
\bibitem{bornletters} M.~Born and A.~Einstein, \emph{The Born-Einstein Letters} (translated by Irene Born), Walker and Company (1971).
\bibitem{shannon}  C.~E.~Shannon,  \emph{A Mathematical Theory of Communication},  The Bell System Technical Journal, Vol. \textbf{27}, pp. 379--423, 623--656, July, October, 1948.
\bibitem{gleick} J.~Gleick,  \emph{The Information},  Pantheon Books (2011).
\bibitem{lloyd} S.~Lloyd, \emph{Programming the Universe: A Quantum Computer Scientist Takes on the Cosmos}, Vintage (2007).
\bibitem{vedral}  V.~Vedral, \emph{Decoding Reality: The Universe as Quantum Information},  Oxford University Press (2010).
\bibitem{dennett} D.~Dennett, \emph{Consciousness Explained},  Back Bay Books (1992).
\end{thebibliography}
\end{document}